\documentclass[journal,comsoc]{IEEEtran}

\usepackage{xspace,amsmath,amssymb,amsfonts,epsfig,syntonly}
\usepackage{cite,bm,color,url,textcomp,empheq,boxedminipage}
\usepackage{algorithmicx,algorithm,algpseudocode}
\usepackage{epstopdf}
\usepackage{empheq}
\usepackage{pifont}

\usepackage{graphicx,graphics}  
\usepackage{multirow,multicol}
\usepackage{psfrag}    
\usepackage{stfloats}
\usepackage{url}

\newtheorem{lemma}{\underline{Lemma}}[section]

\newtheorem{proposition}{\underline{Proposition}}[section]

\newtheorem{remark}{\underline{Remark}}[section]

\DeclareMathOperator*{\argmin}{arg\,min}

\long\def\symbolfootnote[#1]#2{\begingroup
\def\thefootnote{\fnsymbol{footnote}}
\footnote[#1]{#2}\endgroup}

\psfull

\allowdisplaybreaks[4]

\usepackage[T1]{fontenc}
\usepackage{ifpdf}
 \ifpdf
 \else
 \fi

\begin{document}
\title{Optimal Resource Allocation for Wireless Powered Mobile Edge Computing with Dynamic \\ Task Arrivals}

\author{Feng Wang$^\dagger$, Hong Xing$^\ddagger$, and Jie Xu$^{\dagger}$ \\
$^\dagger$School of Information Engineering, Guangdong University of Technology, China\\
$^\ddagger$School of Electrical Engineering and Computer Science, Queen Mary University of London, U.K.\\
E-mail: fengwang13@gdut.edu.cn, h.xing@qmul.ac.uk, jiexu@gdut.edu.cn
\thanks{
This work was supported in part by the National Natural Science Foundation of China (Project No. 61871137 and 61871136), the Project of the Education Department of Guangdong Province (Project No. 2017KZDXM028), and the Natural Science Foundation of Guangdong Province (Project No. 2018A030310537). (J. Xu is the corresponding author.)}

\vspace{-0.9cm}
}

\maketitle

\begin{abstract}
  This paper considers a wireless powered multiuser mobile edge computing (MEC) system, where a multi-antenna access point (AP) employs the radio-frequency (RF) signal based wireless power transfer (WPT) to charge a number of distributed users, and each user utilizes the harvested energy to execute computation tasks via local computing and task offloading. We consider the frequency division multiple access (FDMA) protocol to support simultaneous task offloading from multiple users to the AP. Different from previous works that considered one-shot optimization with static task models, we study the joint computation and wireless resource allocation optimization with dynamic task arrivals over a finite time horizon consisting of multiple slots. Under this setup, our objective is to minimize the system energy consumption including the AP's transmission energy and the MEC server's computing energy over the whole horizon, by jointly optimizing the transmit energy beamforming at the AP, and the local computing and task offloading strategies at the users over different time slots. To characterize the fundamental performance limit of such systems, we focus on the offline optimization by assuming the task and channel information are known {\emph{a-priori}} at the AP. In this case, the energy minimization problem corresponds to a convex optimization problem. Leveraging the Lagrange duality method, we obtain the optimal solution to this problem in a well structure. It is shown that in order to maximize the system energy efficiency, the optimal number of task input-bits at each user and the AP are monotonically increasing over time, and the offloading strategies at different users depend on both the wireless channel conditions and the task load at the AP. Numerical results demonstrate the benefit of the proposed joint-WPT-MEC design over alternative benchmark schemes without such joint design.
\end{abstract}


\vspace{-0.2cm}
\section{Introduction}
Wireless powered mobile edge computing (MEC) has attracted growing research interests to support self-sustainable computation for massive low-power wireless devices, by combining emerging radio-frequency (RF) signal based wireless power transfer (WPT)\cite{Zeng17,Clerckx} and MEC \cite{Sar14,JunZhang17,Xiaowen18} techniques into a joint design\cite{You16,Feng17,Bi18}. By deploying hybrid access points (APs) each with triple roles of energy transmitter, information transceiver, and MEC server, this technique can provide continuous wireless energy supply for end users, such that they can rely on the harvested energy for local computing and computation offloading. As compared with conventional MEC systems with fixed battery supplies at users, the wireless powered MEC can significantly enhance the cost-efficiency and sustainability of future Internet-of-things (IoT) networks, by e.g., avoiding frequent battery replacement at users.

The design of wireless powered MEC systems encounters various new technical challenges due to the involvement of WPT, computation, and communication. In order to optimize the system performance, it is crucial to perform multi-resource allocation to achieve optimal balance between the wireless energy supply from the AP versus the computation and communication energy demand at these users. In the literature, there have been several prior works investigating joint WPT, communication, and computation resource allocation in wireless powered MEC systems, under different system setups with one single user \cite{You16}, multiple users \cite{Feng17,Bi18}, and user cooperation \cite{Hu18,Wu18}, respectively. These works \cite{You16,Feng17,Bi18,Hu18,Wu18} focused on one-shot optimization over a particular time slot by assuming unchanged wireless channels and static task models at users.

In practice, however, due to the randomness of computation traffics, the task arrival rates at users may fluctuate substantially over time and space. On the other hand, due to the randomness of wireless channels, the wireless energy harvested from the AP may also change significantly over time among different users. Therefore, it is an important yet challenging problem to jointly design the WPT at the AP and the computation and communication resource allocation at the users, in order to properly control the users' harvested energy to well match their dynamic computation requirements in an energy-efficient manner. Notice that in the literature, some prior works \cite{Mao16JASC,Mao17TWC,Jie17Cog} investigated joint communication and computation management in MEC systems with dynamic task arrivals over time, while other works (see, e.g., \cite{Rahbar15,Li15} and the references therein) considered dynamic energy management with random energy arrivals for energy harvesting based systems. However, there still lacks a joint design of the WPT at the AP and the communication/computation energy demand at the users for wireless powered MEC systems, by taking into account both energy and task dynamics over time and space. This thus motivates our study in this work.

This paper considers a wireless powered multiuser MEC system, where a multi-antenna AP employs the energy beamforming to charge a number of distributed users with dynamic task arrivals, and each user utilizes the harvested energy to execute the computation tasks via local computing and computation offloading.
Suppose that the downlink WPT from the AP to the users and the uplink computation offloading are operated simultaneously over orthogonal frequency bands, and frequency division multiple access (FDMA) protocol is employed to enable simultaneous task offloading from multiple users to the AP. We focus on joint computation and wireless resource allocation over a finite time horizon consisting of multiple slots. Under this setup, our objective is to minimize the system energy consumption including the AP's transmission energy and the MEC server's computing energy over the whole horizon, by jointly optimizing the transmit energy beamforming at the AP, as well as the local computing and task offloading strategies at the users over different slots.

To further characterize the fundamental performance limit, we assume that the channel state information (CSI) and the task arrival information are perfectly known {\em a-priori} at the AP. Accordingly, we adopt the {\em offline} optimization approach to solve the energy minimization problem.\footnote{Notice that the offline optimization approach has been widely adopted in the energy harvesting based wireless communication \cite{Li15} and the energy management in smart grids \cite{Rahbar15}. The obtained optimal offline solution can serve as the performance upper bound for any practical online designs, and can also inspire practical online designs in the case when the CSI and the task arrival information are casually known. We leave the online optimization for the wireless powered MEC systems to our future work.} Leveraging the Lagrange duality method, we obtain well-structured optimal solution to the joint-WPT-MEC design problem. It is shown that in order to maximize the system energy efficiency, the number of task input-bits at each user and the AP are monotonically increasing over time, and different users' offloading strategies are adapted according to both the wireless channel conditions and the task load at the AP. Numerical results demonstrate the benefit of the proposed joint design over other benchmark schemes without such joint design.


\vspace{-0.3cm}
\section{System Model and Problem Formulation}\label{Sec:System}

\begin{figure}
\vspace{0cm}
  \centering
  \includegraphics[width = 2.8in]{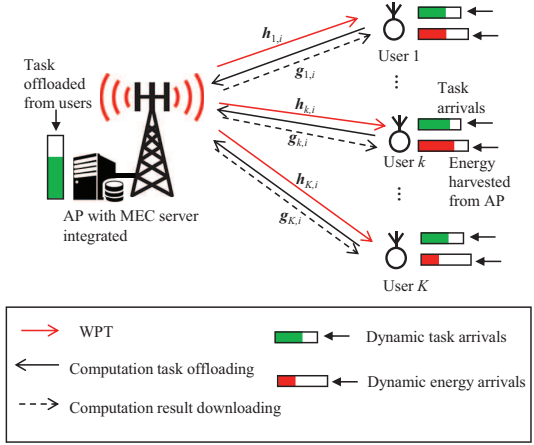}
 \caption{The wireless powered MEC system model.} \label{fig.system-model}
 \vspace{-0.6cm}
\end{figure}

We consider a wireless powered multiuser MEC system as shown in Fig.~\ref{fig.system-model}, where an $M$-antenna AP (integrated with an MEC server) employs WPT to charge a set ${\cal K}\triangleq \{1,\ldots,K\}$ of single-antenna users, and provides cloud-like computation capability to enable their computation-intensive applications. We consider a finite time horizon with duration $T$, which is divided into $N$ slots each with duration $\tau=T/N$. Let ${\cal N}\triangleq \{1,\ldots,N\}$ denote the set of the $N$ slots. For each user $k\in{\cal K}$, the computation tasks arrive at the beginning of each slot $i\in{\cal N}$, with $A_{k,i}\geq 0$ denoting the corresponding number of task input-bits. In order to enable remote computing at the AP, at each slot $i\in{\cal N}$, the users need to offload the computation tasks to the AP and download the computation results of previously executed tasks from the AP. It is assumed that each user is subject to a task completion deadline of $T$, i.e., each user needs to successfully execute their respective tasks before the end of this time horizon. As commonly adopted in the literature\cite{Feng17,You16,Xiaowen18,Bi18}, where the computation results are generally much smaller than the task input-bits, we focus on the time and energy consumed by task offloading from the users to the AP and ignore those consumed by computation result downloading. In the following, we first introduce the task execution at the users via local computing and computation offloading, and then present the WPT and the remote task execution at the AP.
\vspace{-0.4cm}
\subsection{Task Execution and Computation Offloading at Users}

During each slot $i\in{\cal N}$, each user $k\in{\cal K}$ can execute its tasks via local computing and computation offloading. Let $L_{k,i}\geq 0$ and $R_{k,i}\geq 0$ denote the number of task input-bits for user $k$'s local computing and offloading, respectively, which are design variables to be optimized later. In this case, we first have the following {\em task causality constraints}, such that at each slot $i\in{\cal N}$, the cumulative number of task input-bits executed until that slot (i.e., $\sum_{j=1}^i(L_{k,j}+R_{k,j})$) cannot exceed that having already arrived at user $k$ (i.e., $\sum_{j=1}^i A_{k,j}$), i.e.,
\begin{align}\label{eq.Si-user}
&\sum_{j=1}^i A_{k,j} -\sum_{j=1}^i L_{k,j}-\sum_{j=1}^i R_{k,j} \geq 0,\notag  \\
&\quad\quad\quad\quad\quad\quad\quad\quad \forall i\in\{1,\ldots,N-1\}, ~k \in{\cal K}.
\end{align}

In addition, we have the {\em computation deadline constraints} at the users, i.e., each user $k\in{\cal K}$ needs to accomplish the task execution by the end of the last slot $N$, which is expressed as
\begin{align}\label{eq.SN-user}
\sum_{j=1}^{N} A_{k,j} - \sum_{j=1}^N L_{k,j} - \sum_{j=1}^{N} R_{k,j} = 0, ~~\forall k\in {\cal K}.
\end{align}

\subsubsection{Local Computing at Users}
First, we consider the local computing at user $k\in{\cal K}$ for executing $L_{k,i}$ task input-bits at slot $i\in{\cal N}$, for which $C_kL_{k,i}$ CPU cycles in total are required, where $C_k\geq 0$ denotes the number of required central processing unit (CPU) cycles for executing one task input-bit at user $k\in{\cal K}$. Note that the value $C_k$ depends on the type of applications and the CPU architectures of user $k$. In order to maximize the energy efficiency for local computing, each user $k\in{\cal K}$ can apply dynamic voltage and frequency scaling (DVFS) technique by adjusting the CPU frequency as $C_kL_{k,i}/\tau$ during each slot $i\in{\cal N}$\cite{Feng17}. As a result, in slot $i\in{\cal N}$, the energy consumption $E_{k,i}^{\rm loc}$ for local computing at user $k\in{\cal K}$ is
\begin{align}\label{eq.Eki}
E_{k,i}^{\rm loc} = \sum_{n=1}^{C_kL_{k,i}} \zeta_k \left(\frac{C_kL_{k,i}}{\tau} \right)^2 =\frac{\zeta_k C_k^3 L_{k,i}^3}{\tau^2},
\end{align}
where $\zeta_{k}\geq 0$ is the effective switched capacitance coefficient of user $k$'s CPU chip architecture.

\subsubsection{Computation Offloading from Users to AP}
In order to avoid co-channel interference, we adopt the FDMA protocol for the $K$ users to simultaneously offload their task input-bits to the AP for remote server's computing, where each user $k\in{\cal K}$ is allocated with an identical system bandwidth of $B>0$. For each slot $i\in{\cal N}$, let $g_{k,i}>0$ denote the channel power gain from user $k$ to the AP. Assuming the maximal-ratio combining (MRC) based receiver at the AP, the transmission energy consumption at user $k$ for offloading $R_{k,i}$ task input-bits within slot $i\in{\cal N}$ is\cite{Feng17}
\begin{align}
E_{k,i}^{\rm offl} = \frac{\tau\sigma^2}{{g}_{k,i}}(2^{\frac{R_{k,i}}{\tau B}}-1),
\end{align}
where $\sigma^2$ is the additive white Gaussian noise (AWGN) power at the AP receiver.
\vspace{-0.4cm}
\subsection{WPT and Remote Task Execution at AP}
The AP needs to charge the $K$ users via downlink energy beamforming and remotely execute the offloaded computation tasks for the $K$ users.
\subsubsection{WPT from AP to Users} First, consider the downlink energy beamforming from the AP to the $K$ users. At slot $i\in{\cal N }$, let $\bm s_i\in\mathbb{C}^{M\times 1}$ denote the energy-bearing transmit signal and $\bm Q_i\triangleq \mathbb{E}[\bm s_i\bm s_i^H]\in\mathbb{C}^{M\times M}$ the transmit energy covariance matrix of the AP, where ${\mathbb E}[\cdot]$ denotes the expectation operation and the superscript $H$ represents the conjugate transpose of a matrix or vector. For ease of analysis, we consider that the input RF power at each user $k\in{\cal K}$ is within the linear regime of the rectifier.\footnote{In practice, our proposed wireless powered MEC system design is extendable to the scenario with non-linear energy harvesting models at each user, for which the transmit waveform optimization needs to be considered jointly with the energy beamforming design at the AP\cite{Clerckx}.} Then the amount of energy harvested by user $k\in{\cal K}$ at slot $i\in{\cal N}$ is\cite{Zeng17}
\begin{align}\label{eq.EH-model}
E_{k,i}=\tau \eta_k {\rm tr}\left(\bm Q_i\bm h_{k,i}\bm h_{k,i}^H\right),
\end{align}
where $\bm h_{k,i}\in\mathbb{C}^{M\times 1}$ denotes the downlink channel vector from the AP to user $k$ at slot $i$, $0<\eta_k\leq 1$ denotes the constant energy harvesting efficiency of user $k$, and ${\rm tr}(\cdot)$ denotes the trace of a matrix.

\subsubsection{Remote Computing at AP}
Next, we discuss the remote execution at the AP. During slot $i\in{\cal N}$, the AP needs to execute the users' offloaded tasks by its integrated MEC server. Let $L_{0,i}\geq 0$ denote the number of task input-bits computed by the AP within slot $i$. In practice, at each slot $i\in{\cal N}$, the MEC server can only execute the task input-bits offloaded from the users at the previous slots. Hence, at each slot $i\in{\cal N}$, the number of task input-bits cumulatively executed by the AP (i.e., $\sum_{j=1}^i L_{0,j}$) cannot exceed that cumulatively offloaded from all the $K$ users until the previous slot (i.e., $\sum_{j=1}^{i-1} \sum_{k=1}^K R_{k,j}$). As in \eqref{eq.Si-user}, the task causality constraints at the AP are given by
\begin{align}\label{eq.MECi}
\sum_{j=1}^{i-1} \sum_{k=1}^K R_{k,j} - \sum_{j=1}^i L_{0,j} \geq 0,~~\forall i\in \cal{N}.
\end{align}

In addition, the AP needs to complete all the task execution by the last slot $N$. As a result, as in \eqref{eq.SN-user}, the computation deadline constraint for the AP is expressed as
\begin{align}\label{eq.MECN}
\sum_{j=1}^{N-1} \sum_{k=1}^K R_{k,j} - \sum_{j=1}^N L_{0,j} = 0.
\end{align}
Based on \eqref{eq.SN-user}, \eqref{eq.MECi}, and \eqref{eq.MECN}, the number of task input-bits offloaded by user $k$ at slot $N$ must satisfy $R_{k,N}=0$, as there is no time left for the AP to further perform computation before the deadline.

In order to remotely compute the tasks in an energy-efficient manner, the AP employs the DVFS technique by adjusting the CPU frequency as $C_0L_{0,i}/\tau$ during each slot $i\in{\cal N}$, where $C_0\geq 0$ denotes the number of required CPU cycles for executing one task input-bit by the MEC server. In this case, as in \eqref{eq.Eki}, the total amount of computation energy consumption at the AP across the $N$-slot horizon is expressed as
\begin{align}\label{eq.MEC}
E_{\rm MEC} = \sum_{i=1}^N \frac{\zeta_0 C_0^3L^3_{0,i}}{\tau^2},
\end{align}
where $\zeta_0$ is the capacitance coefficient specified by the MEC server's CPU architecture.
\vspace{-0.3cm}
\subsection{Problem Formulation}
Note that each user $k\in{\cal K}$ is powered by the energy beamforming from the AP to achieve self-sustainable computation. In this case, at each slot $i\in{\cal N}$, the cumulatively consumed energy at each user up to that slot cannot exceed that cumulatively harvested from the AP. As a result, we obtain the {\em energy causality constraints} on users' battery storage:
\begin{align}\label{eq.EHi}
\sum_{j=1}^i E_{k,j}^{\rm loc}+\sum_{j=1}^i E_{k,j}^{\rm offl} \leq \sum_{j=1}^{i} E_{k,j}, ~~\forall i\in {\cal N},~k\in{\cal K}.
\end{align}

We are interested in an energy-efficient joint-WPT-MEC design to minimize the system energy consumption subject to the energy/task causality and computation deadline constraints. Specifically, we aim to minimize the energy consumption of the AP, i.e., $\sum_{i=1}^N\tau{\rm tr}(\bm Q_i)+E_{\rm MEC}$, while ensuring the successful task execution before the deadline, by jointly optimizing the energy transmit covariance matrices $\{\bm Q_i\}$, the number of task input-bits $\{L_{0,i}\}$ for remote execution at the AP, and the number of task input-bits $\{L_{k,i}\}$ and $\{R_{k,i}\}$ for users' local computation and computation offloading, respectively. As a result, the energy minimization problem is formulated as
\begin{subequations}\label{eq.prob1}
\begin{align}
&\min_{\{\bm Q_i,L_{0,i},L_{k,i},R_{k,i}\}} ~\sum_{i=1}^{N} \tau{\rm tr}(\bm Q_i) + \sum_{i=1}^N\frac{\zeta_0 C_0^3 L_{0,i}^3}{\tau^2}\\
&\quad\quad {\rm s.t.} ~ \eqref{eq.Si-user},~ \eqref{eq.SN-user}, ~\eqref{eq.MECi}, ~\eqref{eq.MECN},~{\rm and}~\eqref{eq.EHi}\\
&~~ \quad\quad\quad L_{k,i}\geq 0, R_{k,i}\geq 0, L_{0,i}\geq 0,~\forall i\in{\cal N}, k\in{\cal K}\\
&~~\quad\quad\quad \bm Q_i\succeq \bm 0, ~~\forall i\in{\cal N}.
\end{align}
\end{subequations}

In particular, we consider offline optimization for problem \eqref{eq.prob1} by assuming that $\{\bm h_{k,i},g_{k,i}, A_{k,i}\}$ are perfectly known {\em a-priori} at the AP, in order to characterize the fundamental performance upper bound and inspire practical online designs in future work. Under this assumption, problem \eqref{eq.prob1} is a convex optimization problem and can thus be solved by standard convex optimization techniques such as the interior-point method\cite{Boyd2004}. However, to reveal more design insights, we apply the Lagrange duality method to obtain a well-structured optimal solution to problem \eqref{eq.prob1} in Section~\ref{sec:optimal}.

\section{Optimal Solution}\label{sec:optimal}
As problem \eqref{eq.prob1} is convex and satisfies Slater's condition\cite{Boyd2004}, strong duality holds between problem \eqref{eq.prob1} and its dual problem. Let $\mu_{k,i}\geq 0$, $\mu_{k,N}\in\mathbb{R}$, $\nu_i \geq 0$, $\nu_N\in\mathbb{R}$, and $\lambda_{k,i}\geq 0$ denote the Lagrange multipliers associated with the constraints in \eqref{eq.Si-user}, \eqref{eq.SN-user}, \eqref{eq.MECi}, \eqref{eq.MECN}, and \eqref{eq.EHi}, $i\in\{1,\ldots,N-1\}$, $k\in{\cal K}$, respectively. By defining $\bm \lambda \triangleq [\lambda_{1,1}, \ldots,\lambda_{K,N}]^\dagger$, $\bm \mu \triangleq [\mu_{1,1},\ldots,\mu_{K,N}]^\dagger$, and $\bm \nu \triangleq [\nu_1,\ldots,\nu_N]^\dagger$, where the superscript $\dagger$ denotes the transpose of a vector, the partial Lagrangian of problem \eqref{eq.prob1} is given by
\begin{align}\label{eq.Lagrangian}
& {\cal L}(\bm \lambda,\bm \mu,\bm \nu, \{\bm Q_i,L_{0,i},L_{k,i},R_{k,i}\}) \notag\\
&\quad =  \sum_{i=1}^{N} \tau {\rm tr}\left(\bm Q_i \bar{\bm H}_i\right)+\sum_{i=1}^N \left( \frac{\zeta_0 C_0^3 L_{0,i}^3}{\tau^2} + \sum_{j=i}^N\nu_j L_{0,i} \right)\notag\\
&\quad\quad +\sum_{k=1}^K \sum_{i=1}^N \left(\frac{\sum_{j=i}^N\lambda_{k,j}\zeta_kC_k^3L_{k,i}^3}{\tau^2}+\sum_{j=i}^N\mu_{k,j}L_{k,i}\right) \notag \\
&\quad\quad +\sum_{k=1}^K\sum_{i=1}^{N} \Bigg( \frac{\sum_{j=i}^N\lambda_{k,j} \tau\sigma^2}{{g}_{k,i}}\left(2^{\frac{R_{k,i}}{\tau B}}-1\right) \notag \\
& \quad\quad + \sum_{j=i}^N(\mu_{k,j}-\nu_j)R_{k,i}\Bigg)  -\sum_{k=1}^K\sum_{i=1}^{N} \sum_{j=i}^N\mu_{k,j} A_{k,i},
\end{align}
where $\bar{\bm H}_i\triangleq \bm I_M- \sum_{k=1}^K \sum_{j=i}^N\lambda_{k,j}\eta_k\bm h_{k,i}\bm h_{k,i}^H$ for any $i\in{\cal N}$. Accordingly, the dual function is expressed as
\begin{align}\label{eq.dual-func}
&{\cal G}( \bm \lambda,\bm \mu,\bm\nu )\notag\\
&= \min_{\substack{\{\bm Q_i\succeq \bm 0, L_{0,i}\geq 0\}\\
\{L_{k,i}\geq 0,R_{k,i}\geq 0\}}} {\cal L}(\bm \lambda,\bm \mu,\bm\nu,\{\bm Q_i,L_{0,i},L_{k,i},R_{k,i}\}).
\end{align}
In order to obtain the dual problem of problem \eqref{eq.prob1}, we first establish the following lemma.
\begin{lemma}\label{lem.bound}
In order for the dual function ${\cal G}(\bm \lambda,\bm \mu,\bm \nu)$ to be lower bounded from below, it must hold that
\begin{subequations}\label{eq.bound}
\begin{align}
&\bar{\bm H}_i \succeq \bm 0,~\forall i\in{\cal N},\\
&\sum_{j=i}^N\lambda_{k,j}>0,~\forall i\in{\cal N},~ k\in{\cal K}.
\end{align}
\end{subequations}
\end{lemma}

\begin{IEEEproof}
See Appendix A.
\end{IEEEproof}

Based on Lemma \ref{lem.bound}, the dual problem of problem \eqref{eq.prob1} is then expressed as
\begin{subequations}\label{eq.prob1_dual}
\begin{align}
&\max_{\bm\lambda,\bm \mu,\bm \nu } ~{\cal G}(\bm\lambda,\bm \mu,\bm \nu)\\
&~~{\rm s.t.} ~~\bar{\bm H}_i \succeq \bm 0,~~\forall i\in{\cal N}\\
&~~~~~~~ \sum_{j=i}^N \lambda_{k,j}> 0,~\lambda_{k,i} \geq 0,~\forall k\in{\cal K}, i\in{\cal N}\\
&~~~~~~~ \mu_{k,i}\geq 0, \nu_i \geq 0, ~\forall k\in{\cal K},i\in\{1,\ldots,N-1\}.
\end{align}
\end{subequations}
Denote $\cal X$ as the feasible solution set of $(\bm\lambda,\bm \mu,\bm \nu)$ for problem \eqref{eq.prob1_dual}. Then, we solve problem \eqref{eq.prob1} optimally by solving its dual problem \eqref{eq.prob1_dual} equivalently.

\subsection{Evaluating Dual Function ${\cal G}(\bm \lambda,\bm \mu,\bm \nu)$} 
Under any given $(\bm \lambda,\bm \mu,\bm \nu)\in{\cal X}$, by removing the irrelevant constant terms in \eqref{eq.Lagrangian}, the optimization problem in \eqref{eq.dual-func} can be decomposed into the following $(2N+2NK)$ independent subproblems for different time slots and users.
\begin{align}
&\min_{\bm Q_i\succeq \bm 0} ~~{\rm tr}\left(\bm Q_i\bar{\bm H}_i\right),~~\forall i\in{\cal N}\label{eq.subp1}\\
&\min_{L_{0,i}\geq 0}~~ \frac{\zeta_0 C_0^3 L_{0,i}^3}{\tau^2} + \sum_{j=i}^N\nu_j L_{0,i},~~\forall i\in{\cal N} \label{eq.subp1-2}\\
&\min_{L_{k,i}\geq 0} \frac{\sum_{j=i}^N\lambda_{k,j}\zeta_k C_k^3L_{k,i}^3}{\tau^2}+\sum_{j=i}^N\mu_{k,j}L_{k,i}, \notag \\
&\quad\quad\quad\quad\quad \quad\quad\quad\quad\quad\quad\quad\quad\quad\quad \forall i\in{\cal N},k\in{\cal K}\label{eq.subp2}\\
&\min_{R_{k,i}\geq 0} \frac{\sum_{j=i}^N\lambda_{k,j} \tau \sigma^2}{{g}_{k,i}}(2^{\frac{R_{k,i}}{\tau B}}-1) +\sum_{j=i}^N(\mu_{k,j}-\nu_j)R_{k,i}, \notag \\
&\quad\quad\quad\quad\quad\quad\quad\quad\quad\quad\quad\quad\quad\quad\quad \forall i\in{\cal N},k\in{\cal K} \label{eq.subp3}
\end{align}
Let $\bm Q_i^*$, $L^*_{0,i}$, $L_{k,i}^*$, and $R_{k,i}^*$, $k\in{\cal K}$, $i\in{\cal N}$, denote the optimal solutions to the subproblems in \eqref{eq.subp1}, \eqref{eq.subp1-2}, \eqref{eq.subp2}, and \eqref{eq.subp3}, respectively. We establish the following lemmas.
\begin{lemma}\label{lem.sub1}
 The optimal solution $\{\bm Q_i^*\}$ to problem \eqref{eq.subp1} is given by $\bm Q_i^* \in {\rm Null}({\bar{\bm H}})$, $i\in{\cal N}$, where ${\rm Null}(\cdot)$ denotes the null space of a matrix.
\end{lemma}
\begin{IEEEproof}
As the matrix $\bar{\bm H}_{i}$ is semidefinite positive in problem \eqref{eq.subp1}, with $\bm Q_i\succeq \bm 0$ , it follows that the minimal value of ${\rm tr}(\bm Q_i\bar{\bm H}_{i})$ is zero. Therefore, the optimal $\{\bm Q_i^*\}$ to problem \eqref{eq.subp1} must satisfy $\bm Q_i^*\in{\rm Null}(\bar{\bm H}_{i})$, $\forall i\in{\cal N}$.
\end{IEEEproof}

\begin{lemma}\label{lem.sub1-2}
 The optimal solution $\{L^*_{0,i}\}$ to problem \eqref{eq.subp1-2} is given by
\begin{align}\label{eq.dual-sub1-2}
L_{0,i}^* = \sqrt{\frac{ \tau^2 \left[-\sum_{j=i}^N\nu_j\right]^+}{3\zeta_0 C_0^3}},~~\forall i\in{\cal N},
\end{align}
where $[x]^+=\max(x,0)$.
\end{lemma}
\begin{IEEEproof}
See Appendix B.
\end{IEEEproof}

\begin{lemma}\label{lem.sub2}
 The optimal solution $\{L^*_{k,i}\}$ to problem \eqref{eq.subp2} is given by
\begin{align}\label{eq.dual-sub2}
L_{k,i}^* = \sqrt{\frac{\tau^2\left[-\sum_{j=i}^N \mu_{k,j}\right]^+}{3\sum_{j=i}^N \lambda_{k,j}\zeta_kC^3_k}},~~\forall i\in{\cal N},k\in{\cal K}.
\end{align}
\end{lemma}

\begin{lemma}\label{lem.sub3}
 The optimal solution $\{R_{k,i}^*\}$ to problem \eqref{eq.subp3} is given by, $\forall i\in{\cal N},k\in{\cal K}$,
\begin{align}\label{eq.dual-sub3}
&R_{k,i}^* = \begin{cases}
\left[\tau B\log_2\left(\frac{\sum_{j=i}^N\left(\nu_{j}-\mu_{k,j}\right)B{g}_{k,i}}{\sum_{j=i}^N\lambda_{k,j}\tau\sigma^2\ln2}\right)\right]^+,\\
~~~~~~~~~~~~~~{\rm if}~\sum_{j=i}^N( \nu_{j}-\mu_{k,j}) >0,\\
0,\\
~~~~~~~~~~~~~~{\rm if}~\sum_{j=i}^N(\nu_{j}-\mu_{k,j}) \leq 0.
\end{cases}
\end{align}
\end{lemma}
Note that Lemmas \ref{lem.sub2} and \ref{lem.sub3} can be are similarly proved as for Lemma \ref{lem.sub1-2}, and thus we omit the proofs for brevity. Based on Lemma \ref{lem.sub1}, it follows that $\bm Q_i^*$ is not unique if $\bar{\bm H}_i$ is rank deficient. For simplicity, we choose $\bm Q_i^*=\bm 0$, $i\in{\cal N}$, for problem \eqref{eq.subp1} without loss of optimality for obtaining the dual function only. With \eqref{eq.dual-sub1-2}--\eqref{eq.dual-sub3} and $\bm Q_i^*=\bm 0$, $\forall i\in{\cal N}$, ${\cal G}(\bm \lambda,\bm \mu,\bm \nu)$ can be obtained under any given set $(\bm \lambda,\bm \mu,\bm \nu)\in{\cal X}$.

\subsection{Obtaining Optimal Dual $(\bm \lambda^{\rm opt},\bm \mu^{\rm opt},\bm \nu^{\rm opt})$}
Next, we maximize ${\cal G}(\bm \lambda,\bm \mu,\bm \nu)$ over $(\bm \lambda,\bm \mu,\bm \nu)$ to solve the dual problem \eqref{eq.prob1_dual}. Note that the dual function ${\cal G}(\bm \lambda,\bm \mu,\bm \nu)$ is always concave but not necessarily differentiable. Therefore, problem \eqref{eq.prob1_dual} is convex, and can thus be solved by subgradient based methods such as the ellipsoid method \cite{Boyd2004}, in which the subgradient of ${\cal G}(\bm \lambda,\bm \mu, \bm \nu)$ is $\Big[
\frac{\zeta_1C_1^3L_{1,1}^{*3}}{\tau^2}+\frac{\tau\sigma^2}{g_{1,1}}(2^{\frac{R^*_{1,1}}{\tau B}}-1),\ldots,\sum_{j=1}^N \frac{\zeta_KC_K^3L_{K,j}^{*3}}{\tau^2}+\sum_{j=1}^N\frac{\tau\sigma^2}{g_{K,j}}(2^{\frac{R^*_{K,j}}{\tau B}}-1), L^*_{1,1}+ R^*_{1,1}-A_{1,1},\ldots,\sum_{j=1}^N L^*_{K,j}+ \sum_{j=1}^NR^*_{K,j}-\sum_{j=1}^N A_{K,j},0,L^*_{0,2}-\sum_{k=1}^K R^*_{K,1},\ldots,\sum_{j=1}^NL^*_{0,j}-\sum_{j=1}^{N-1}\sum_{k=1}^K R^*_{k,j}\Big]^\dagger\in \mathbb{C}^{(2NK+N)\times 1}$ with respect to $(\bm \lambda,\bm \mu,\bm \nu)$. Let $(\bm \lambda^{\rm opt},\bm \mu^{\rm opt},\bm \nu^{\rm opt})$ denote the obtained optimal dual solution to problem~\eqref{eq.prob1_dual}.

\subsection{Finding Optimal Primal $\{\bm Q_i^{\rm opt},L_{0,i}^{\rm opt},L_{k,i}^{\rm opt},R_{k,i}^{\rm opt}\}$}
With $(\bm \lambda^{\rm opt},\bm \mu^{\rm opt},\bm \nu^{\rm opt})$ obtained, it remains to find the optimal primal solution to problem \eqref{eq.prob1}. Since $\{L_{0,i}^*\}$, $\{L_{k,i}^*\}$ and $\{R_{k,i}^*\}$ are the unique optimal solution to problems \eqref{eq.subp1-2}, \eqref{eq.subp2}, and \eqref{eq.subp3}, respectively, the optimal $\{L_{0,i}^{\rm opt}\}$, $\{L_{k,i}^{\rm opt}\}$, and $\{R_{k,i}^{\rm opt}\}$ to problem \eqref{eq.prob1} can be directly obtained by replacing $(\bm \lambda,\bm \mu,\bm \nu)$ with the optimal dual $(\bm \lambda^{\rm opt}, \bm \mu^{\rm opt},\bm \nu^{\rm opt})$ in \eqref{eq.dual-sub1-2}, \eqref{eq.dual-sub2}, and \eqref{eq.dual-sub3}, respectively.

On the other hand, the optimal solution to problem \eqref{eq.prob1} cannot be obtained from Lemma \ref{lem.sub1} alone, and $\bm Q_i^*=\bm 0$, $i\in{\cal N}$, are even not feasible for problem \eqref{eq.prob1}. Therefore, an additional procedure is required to obtain the optimal $\{\bm Q_i^{\rm opt}\}$. With $\{L_{0,i}^{\rm opt},L_{k,i}^{\rm opt},R_{k,i}^{\rm opt}\}$ obtained, the optimal $\{\bm Q_i^{\rm opt}\}$ to problem \eqref{eq.prob1} can be obtained by solving the following semidefinite program (SDP) via convex solvers (e.g., CVX toolbox \cite{Boyd2004}):
\begin{subequations}\label{eq.prob-Q}
\begin{align}
 & \{\bm Q_i^{\rm opt}\}\triangleq \argmin_{\{\bm Q_i\succeq \bm 0\}} ~\sum_{i=1}^{N} \tau {\rm tr}(\bm Q_i)\\
&\quad \quad {\rm s.t.} ~ \sum_{j=1}^i\frac{\zeta_kC^3_k(L^{\rm opt}_{k,j})^3}{\tau^2}+\sum_{j=1}^i \frac{\tau \sigma^2}{{g}_{k,j}}(2^{\frac{R^{\rm opt}_{k,j}}{\tau B}}-1)\notag \\
&\quad\quad \quad~~
\leq \sum_{j=1}^{N} \tau\eta_k{\rm tr}(\bm Q_j\bm h_{k,j}\bm h_{k,j}^H),~ \forall k\in{\cal K}, i\in{\cal N}.
\end{align}
\end{subequations}
By combining $\{\bm Q_i^{\rm opt}\}$ together with $\{ L_{0,i}^{\rm opt}, L_{k,i}^{\rm opt},R_{k,i}^{\rm opt}\}$, we finally obtain the optimal solution to problem \eqref{eq.prob1}. In summary, Algorithm 1 for solving problem \eqref{eq.prob1} is presented in Table~I.

\begin{table}[htp]\label{Alg1}
\begin{center}
\caption{Algorithm 1 for Optimally Solving Problem \eqref{eq.prob1}}
\hrule
\begin{itemize}
\item[a)] {\bf Initialize} $(\bm \lambda,\bm \mu,\bm \nu)$ with $\lambda_{k,i}\geq 0$, $\mu_{k,i}\geq 0$, $\nu_k\geq 0$, $\bar{\bm H}_i \succeq \bm 0$, and $\sum_{j=1}^N \lambda_{k,j}>0$, $\forall k\in{\cal K}$, $i\in{\cal N}$.
\item[b)] {\bf Repeat:}
    \begin{itemize}
    \item[1)]  Obtain $\{L_{0,i}^*\}$, $\{L_{k,i}^*\}$, and $\{R_{k,i}^*\}$ under given $(\bm \lambda,\bm \mu,\bm \nu)$ according to \eqref{eq.dual-sub1-2}, \eqref{eq.dual-sub2}, and \eqref{eq.dual-sub3}, respectively;
    \item[2)]  Update $(\bm \lambda,\bm \mu,\bm \nu)$ based on the ellipsoid method\cite{Boyd2004}, by using the fact that the subgradient of ${\cal G}(\bm \lambda,\bm \mu,\bm \nu)$ is $\Big[
        \frac{\zeta_1C_1^3L_{1,1}^{*3}}{\tau^2}+\frac{\tau\sigma^2}{g_{1,1}}(2^{\frac{R^*_{1,1}}{\tau B}}-1),\ldots,\sum_{j=1}^N \frac{\zeta_KC_K^3L_{K,j}^{*3}}{\tau^2}+\sum_{j=1}^N\frac{\tau\sigma^2}{g_{K,j}}(2^{\frac{R^*_{K,j}}{\tau B}}-1), L^*_{1,1}+ R^*_{1,1}-A_{1,1},\ldots,\sum_{j=1}^N L^*_{K,j}+ \sum_{j=1}^N R^*_{K,j}-\sum_{j=1}^N A_{K,j},0,L^*_{0,2}-\sum_{k=1}^K R^*_{K,1},\ldots,\sum_{j=1}^NL^*_{0,j}-\sum_{j=1}^{N-1}\sum_{k=1}^K R^*_{k,j}\Big]^\dagger\in \mathbb{C}^{(2NK+N)\times 1}$ with respect to $(\bm \lambda,\bm \mu.\bm \nu)$.
    \end{itemize}
\item[c)] {\bf Until} the dual variables $(\bm \lambda,\bm \mu,\bm \nu)$ converge within the prescribed accuracy.
\item[d)] {\bf Set} $(\bm \lambda^{\rm opt},\bm \mu^{\rm opt},\bm \nu^{\rm opt})\leftarrow (\bm \lambda,\bm \mu,\bm \nu)$.
\item[e)] {\bf Output}:
Obtain $\{ L_{0,i}^{\rm opt}\}$, $\{ L^{\rm opt}_{k,i}\}$, and $\{R^{\rm opt}_{k,i}\}$ under the dual optimal $(\bm \lambda^{\rm opt},\bm \mu^{\rm opt},\bm \nu^{\rm opt})$ according to \eqref{eq.dual-sub1-2}, \eqref{eq.dual-sub2}, and \eqref{eq.dual-sub3}, respectively, and compute $\{\bm Q_i^{\rm opt}\}$ by solving the SDP in \eqref{eq.prob-Q}.
\end{itemize}
\hrule \label{algorithm:new0}
\end{center}
\end{table}

To gain essential design insights, based on Lemmas \ref{lem.sub1-2} and~\ref{lem.sub2}, we have Proposition \ref{prop1} for the optimal executed task input-bits allocated to the users' local computing and the AP's remote computing as follows.
\begin{proposition}\label{prop1}
For any user $k\in{\cal K}$ and the AP ($k=0$), the optimal number of executed task input-bits $\{L_{k,i}^{\rm opt}\}$ is monotonically increasing over time, i.e., \begin{align}
L^{\rm opt}_{k,1}\leq\ldots \leq L_{k,N}^{\rm opt}, ~~\forall k\in{\cal K}\cup\{0\}.
\end{align}
\end{proposition}
\begin{IEEEproof}
See Appendix C.
\end{IEEEproof}

\begin{remark}
Proposition \ref{prop1} indicates the monotonically increasing feature of the optimal number of executed task input-bits for both the users and the AP. This can be intuitively understood as follows. Notice that the computation energy consumption functions in \eqref{eq.Eki} and \eqref{eq.MEC} for each slot are convex with respect to the number of task input-bits. Therefore, in order to reduce the energy consumption, it is desirable for the users and the AP to evenly distribute the computation tasks as far as possible. Due to the task causality constraints in \eqref{eq.Si-user} and \eqref{eq.MECi}, it is evident that the computation load at each of the users and the AP should monotonically increase over time, as more tasks will be accumulated over time. This feature is reminiscent of the monotonically increasing power allocation in energy harvesting based wireless communications due to the energy causality constraints (see, e.g., \cite{Li15})..
\end{remark}


\begin{remark}
Based on \eqref{eq.dual-sub3} in Lemma \ref{lem.sub3}, in slot $i\in{\cal N}$, the optimal number of offloaded task input-bits $R_{k,i}^{\rm opt}$ for each user $k\in{\cal K}$ should be adapted according to the task load at the AP. Specifically, under the case when the channels for both WPT and computation offloading remain unchanged between user $k\in{\cal K}$ and the AP, it holds that $R_{k,i+1}\geq R_{k,i}$ if $\nu_i \leq \mu_{k,i}$, $\forall i\in\{1,\ldots,N-2\}$. This implies that, when the penalty in energy consumption for the AP to violate the task causality constraint at slot $i$ is smaller than that for user $k$, user $k$ should offload more task input-bits at the next slot $i+1$, $\forall i\in\{1,\ldots,N-2\}$. In addition, it is intuitively expected that $R_{k,i}^{\rm opt}$ is propositional to the offloading channel power gain $g_{k,i}$ from user $k$ to the AP at slot $i\in{\cal N}$.
\end{remark}

%


\vspace{-0.2cm}
\section{Numerical Results}
In this section, we provide numerical results to evaluate the performance of the proposed algorithm, where $M=4$ and $N=10$ are set with each slot duration being $\tau=0.1$ second. For comparison, we consider the following four benchmark schemes.

\begin{itemize}
\item {\em Local computing only:} Each user $k\in{\cal K}$ accomplishes its computation tasks by only local computing. This scheme corresponds to solving problem \eqref{eq.prob1} by setting $R_{k,i}=0$ and $L_{0,i}=0$, $\forall i\in{\cal N}$, $k\in{\cal K}$. 
\item {\em Full offloading:} Each user $k\in{\cal K}$ accomplishes its computation tasks by fully offloading them to the AP. This scheme corresponds to solving problem \eqref{eq.prob1} by setting $L_{k,i}=0$, $\forall i\in{\cal N}$, $k\in{\cal K}$.
\item {\em Myopic design:} In each slot $i\in{\cal N}$, both user $k\in{\cal K}$ and the AP accomplish their task input-bits of $A_{k,i}$ and $\sum_{k=1}^K R_{k,i-1}$, respectively, i.e., we have $A_{k,i}-L_{k,i}-R_{k,i}=0$ and $L_{0,i}-\sum_{k=1}^K R_{k,i-1}=0$, $\forall i\in{\cal N}$, $k\in{\cal K}$. In this scheme, the system energy minimization can be implemented independently over each individual slot, as investigated in \cite{Feng17}.
\item {\em Separate WPT-MEC design:} This scheme separately designs the energy beamforming for WPT and the users' computation offloading for MEC. First, the sum-energy consumption of the $K$ users is minimized subject to their individual computation latency constraints\cite{JunZhang17}. The AP then designs its energy beamforming with energy minimization under the given energy demands at users.
\end{itemize}

In the simulation, the system parameters are set as follows, unless stated otherwise. We set $\eta_k=0.3$, $C_0=C_k=10^3$ CPU cycles/bit, $\zeta_k=10^{-28}$, $\zeta_k=10^{-29}$, $\forall k\in{\cal K}$, the receiver noise power $\sigma^2=10^{-9}$ Watt, and the system bandwidth for offloading $B=2$ MHz. We also consider a distance-dependent Rayleigh fading channel model \cite{Feng17} with the channel power gain at a unit of reference distance set as $-32$dB. 
The number $A_{k,i}$ of task input-bits for user $k\in{\cal K}$ at slot $i\in{\cal N}$ is set as a random variable following the uniform distribution of $A_{k,i}\sim U(A_{\min},A_{\max})$ with $A_{\min}=10^5$ and $A_{\max} = 10^6$ input-bits unless stated otherwise. The numerical results are obtained by averaging over $500$ randomized channel realizations and randomized task realizations.

\begin{figure}
\vspace{-0.2cm}
  \centering
  \includegraphics[width = 2.8in]{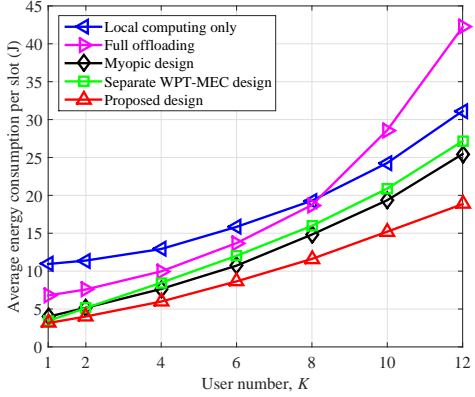}
 \caption{The average energy consumption per slot at the AP versus the user number $K$.} \label{fig.vsK}
 \vspace{-0.2cm}
\end{figure}

Fig. \ref{fig.vsK} shows the average energy consumption per slot at the AP versus the user number $K$, where the distances between the AP and the $K$ users are identical with $d_k= 4$ meters, $\forall k\in{\cal K}$. It is observed that the proposed design achieves the lowest average energy consumption among all the five schemes, and its gain over the benchmark schemes becomes more significant when $K$ increases. At small value of $K$ (e.g., $K\leq 4$), the benchmark myopic-design and separate-WPT-MEC-design schemes are observed to achieve near-optimal performance close to that achieved by the proposed design. It is also observed that the separate-WPT-MEC-design performs inferior to the myopic-design scheme at large $K$ values. This validates the importance of the proposed joint design for energy saving in multiuser scenarios. Both the local-computing-only and full-offloading schemes are observed to perform inferior to the other two benchmark schemes. When $K\leq 8$, the full-offloading scheme outperforms the local-computing-only scheme, and the reverse is true when $K$ becomes larger. This is due to the fact that the energy consumption for offloading increases faster (exponentially) than that for local computing (cubically).

\begin{figure}
\vspace{-0.2cm}
  \centering
  \includegraphics[width = 2.8in]{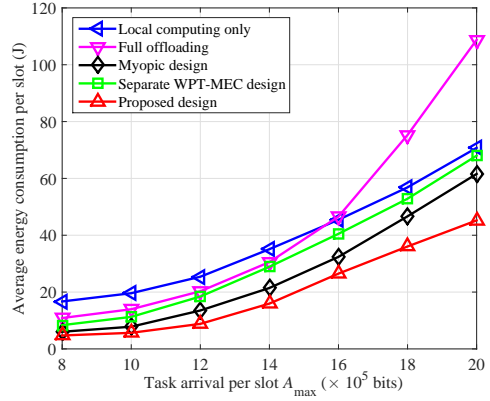}
 \caption{The average energy consumption per slot at the AP versus the maximum number of task input-bits $A_{\max}$.} \label{fig.vsAmax}
 \vspace{-0.2cm}
\end{figure}

Fig. \ref{fig.vsAmax} shows the average energy consumption at the AP per slot versus the maximum number of task input-bits $A_{\max}$, where $K=6$ and $d_k=4$ meters, $\forall k\in{\cal K}$. It is observed that the proposed design outperforms the other benchmark schemes. As $A_{\max}$ increases, the performance gain achieved by the proposed design becomes more substantial. The myopic-design scheme is observed to achieve a near optimal performance close to the proposed design at small $A_{\max}$ values (e.g., $A_{\max}=8\times 10^5$). Similar performance trends for the benchmark schemes are observed as in Fig. \ref{fig.vsK}.

\vspace{-0.2cm}
\section{Conclusion}
This paper studied the optimal resource allocation for a wireless powered multiuser MEC system to minimize the energy consumption at the AP subject to the computation latency and energy harvesting constraints at users, by jointly optimizing the AP's energy beamforming for WPT and remote execution, as well as each user's local computing/offloading. Using the Lagrange duality method, we developed an efficient algorithm to obtain a well-structured optimal solution, where the optimal number of task input-bits at the AP and users is monotonically increasing over time. Numerical results demonstrated the merit of our proposed joint-WPT-MEC design compared to other benchmark schemes without such joint design. In future work, we will extend the results to other setups when the energy/task arrival information and the CSI are only causally known or even a-priori unknown at the AP. For such cases, our offline optimization will serve as a performance upper bound, and the obtained structured solution may motivate online algorithms. In general, how to obtain optimal solution for such new scenarios will be a challenging problem that requires further investigation.


\vspace{-0.2cm}
\appendix
\vspace{-0.2cm}
\subsection{Proof of Lemma \ref{lem.bound}}
First, we prove (\ref{eq.bound}a). Suppose that the matrix $\bar{\bm H}_i$ is not semi-definite positive for some $i\in\{0,\ldots,N-1\}$. In this case, by letting $\bm Q_i=\tau \bm x \bm x^H$, $\tau \rightarrow \infty $, where $ \bm x^H \bar{\bm H}_i\bm x <0$ and $\bm x\in\mathbb{C}^{M\times 1}$, it can be shown from \eqref{eq.Lagrangian} that ${\cal L}(\cdot)\rightarrow -\infty$. Thus the matrix $\bar{\bm H}_i\succeq \bm 0$ must hold for all $i\in{\cal N}$ for ${\cal G}(\bm \lambda,\bm \mu,\bm \nu)$ to be bounded from below. Next, we consider (\ref{eq.bound}b). Suppose that $\sum_{j=i}^N\lambda_j=0$ for some $k\in{\cal K}$ and $i\in{\cal N}$. In this case, by letting $\sum_{j=i}^N\mu_{k,j}<0$, $R_{k,i}=0$, and $L_{k,i}\rightarrow \infty$, it can be shown from \eqref{eq.Lagrangian} that ${\cal L}(\cdot)\rightarrow -\infty$. Thus, the fact $\sum_{j=i}^N\lambda_{k,j}=0$ cannot be true for ${\cal G}(\bm \lambda,\bm \mu,\bm \nu)$ to be bounded from below. Lemma~\ref{lem.bound} is now proved.


\vspace{-0.2cm}
\subsection{Proof of Lemma \ref{lem.sub1-2}}
Introducing the Lagrange multiplier $\theta \geq 0$ associated with the constraint $L_{0,i}\geq 0$, the Lagrangian of problem \eqref{eq.subp1-2} is
${\cal L}_0(L_{0,i},\theta) = \frac{\zeta_0 C_0^3 L_{0,i}^3}{N\tau^2} + \sum_{j=i}^N\nu_j L_{0,i} - \theta L_{0,i}$
and the dual problem of problem \eqref{eq.subp1-2} is then
\begin{align}\label{eq.subp1-2-dual}
\max_{\theta\geq 0}\min_{L_{0,i}\geq 0}{\cal L}_0(L_{0,i},\theta)
\end{align}
Note that problem \eqref{eq.subp1-2} is convex and satisfies Slater's condition. Therefore, the strong duality holds between problems \eqref{eq.subp1-2} and \eqref{eq.subp1-2-dual}. Let $L_{0,i}^*$ and $\theta^*$ be the optimal solutions for problems \eqref{eq.subp1-2} and \eqref{eq.subp1-2-dual}, respectively. The Karush-Kuhn-Tucker (KKT) optimal conditions are then given by
\begin{subequations}\label{eq.subp1-2-dual-KKT}
\begin{align}
&\frac{2\zeta_0C_0^3(L_{0,i}^*)^2}{N\tau^2}+\sum_{j=i}^N \nu_j -\theta^* = 0\\
&\theta^* L^*_{0,i}=0,~~L^*_{0,i}\geq 0,~~\theta^* \geq 0,
\end{align}
\end{subequations}
where (\ref{eq.subp1-2-dual-KKT}a) denotes that the gradient of ${\cal L}_0(L_{0,i},\theta)$ must vanish at $L_{0,i}^*$ and the first equality in (\ref{eq.subp1-2-dual-KKT}b) is the complementary slackness condition. Note that $\theta^*$ acts as a slack variable in (\ref{eq.subp1-2-dual-KKT}a) and it can then be eliminated. Together with $L^*_{0,i}\geq 0$, we obtain $
L_{0,i}^* = \sqrt{\frac{N\tau^2 \left[-\sum_{j=i}^N\nu_j\right]^+}{3\zeta_0 C_0^3}}$, $\forall i\in{\cal N}$.

\vspace{-0.4cm}
\subsection{Proof of Proposition \ref{prop1}}
First, consider the case with $k=0$. Based on Lemma~\ref{lem.sub1-2}, $\forall i\in\{1,\ldots,N-1\}$, we have
\begin{align}
L_{0,i+1}^{\rm opt} &= \sqrt{\frac{N\tau^2 \left[-\sum_{j=i+1}^N\nu^{\rm opt}_{j}\right]^+}{3\zeta_0 C_0^3}}\\
&\geq \sqrt{\frac{N\tau^2 \left[-\sum_{j=i}^N\nu^{\rm opt}_{j}\right]^+}{3\zeta_0 C_0^3}} = L_{0,i}^{\rm opt},
\end{align}
where the inequality follows from the fact that $\nu^{\rm opt}_{i}\geq 0$ for any $i\in\{1,\ldots,N-1\}$. It thus holds that $L^{\rm opt}_{0,1}\leq\ldots \leq L_{0,N}^{\rm opt}$.

Next, we consider the case with $k\in{\cal K}$. Similarly, from Lemma~\ref{lem.sub2}, $\forall i\in\{1,\ldots,N-1\}$, it follows that
\begin{align}
L^{\rm opt}_{k,i+1} &= \sqrt{\frac{\tau^2\left[-\sum_{j=i+1}^N \mu_{k,j}^{\rm opt}\right]^+}{3\sum_{j=i+1}^N \lambda^{\rm opt}_{k,j}\zeta_kC^3_k}}\\
&\geq \sqrt{\frac{\tau^2\left[-\sum_{j=i}^N \mu_{k,j}^{\rm opt}\right]^+}{3\sum_{j=i}^N \lambda^{\rm opt}_{k,j}\zeta_kC^3_k}} = L^{\rm opt}_{k,i},~~\forall k\in{\cal K},
\end{align}
where the inequality follows from the fact that both $\mu_{k,i}^{\rm opt}\geq 0$ and $\lambda_{k,i}^{\rm opt}\geq 0$ hold for any $i\in\{1,\ldots,N-1\}$. Therefore, we have $L^{\rm opt}_{k,1}\leq\ldots \leq L_{k,N}^{\rm opt}$ for all $k\in{\cal K}\cup\{0\}$.


\vspace{-0.2cm}
\begin{thebibliography}{99}

\bibitem{Zeng17}
Y. Zeng, B. Clerckx, and R. Zhang, ``Communications and signals design for wireless power transmission,'' {\em IEEE Trans. Commun.}, vol. 65, no. 5, pp. 2264--2290, May 2017.

\bibitem{Clerckx}
B. Clerckx, R. Zhang, R. Schober, D. W. K. Ng, D. I. Kim, and H. V. Poor, ``Fundamentals of wireless information and power transfer: From RF energy harvester models to signal and system designs,'' to appear in {\em IEEE J. Sel. Areas Commun.}, 2018.

\bibitem{Sar14}
S. Barbarossa, S. Sardellitti, and P. D. Lorenzo, ``Communicating while computing: Distributed mobile cloud computing over 5G heterogeneous networks,'' {\em IEEE Signal Process. Mag.}, vol. 31, no. 6, pp. 45--55, Nov. 2014.

\bibitem{JunZhang17}
Y. Mao, C. You, J. Zhang, K. Huang, and K. B. Letaief, ``A survey on mobile edge computing: The communication perspective,'' {\em IEEE Commun. Surveys Tuts.,} vol. 19, no. 4, pp. 2322--2358, 4th Quar. 2017.

\bibitem{Xiaowen18}
X. Cao, F. Wang, J. Xu, R. Zhang, and S. Cui, ``Joint computation and communication cooperation for energy-efficient mobile edge computing,'' to appear in {\em IEEE Internet Thing J.}, 2018.

\bibitem{You16}
C. You, K. Huang, and H. Chae, ``Energy efficient mobile cloud
computing powered by wireless energy transfer,'' {\em IEEE J. Sel. Areas Commun.}, vol. 34, no. 5, pp. 1757--1771, May 2016.

\bibitem{Feng17}
F. Wang, J. Xu, X. Wang, and S. Cui, ``Joint offloading and computing optimization in wireless powered mobile-edge computing systems,'' {\em IEEE Trans. Wireless Commun.}, vol. 17, no. 3, pp. 1784--1797, Mar. 2018.

\bibitem{Bi18}
S. Bi and Y. J. A. Zhang, ``Computation rate maximization for wireless powered mobile-edge computing with binary computation offloading,'' {\em IEEE Trans. Wireless Commun.}, vol. 17, no. 6, pp. 4177--4190, Jun. 2018.

\bibitem{Hu18}
X. Hu, K.-K. Wong, and K. Yang, ``Wireless powered cooperation-assisted mobile edge computing,'' {\em IEEE Trans. Wireless Commun.}, vol. 17, no. 4, pp. 2375--2388, Apr. 2018.

\bibitem{Wu18}
D. Wu, F. Wang, X. Cao, and J. Xu, ``Wireless powered user cooperative computation in mobile edge computing,'' in {\em Proc. IEEE Globecom Workshops-WEHCN}, Abu Dhabi, UAE, Dec. 2018, pp. 1--7.








\bibitem{Mao16JASC}
 Y. Mao, J. Zhang, and K. B. Letaief, ``Dynamic computation offloading for mobile-edge computing with energy harvesting devices,'' {\em IEEE J. Sel. Areas Commun.}, vol. 34, no. 12, pp. 3590--3605, Dec. 2016

\bibitem{Mao17TWC}
Y. Mao, J. Zhang, S. H. Song, and K. B. Letaief, ``Stochastic joint radio and computational resource management for multi-user mobile-edge computing systems,'' {\em IEEE Trans. Wireless Commun.}, vol. 16, no. 9, pp. 5994--6009, Sep. 2017.

\bibitem{Jie17Cog}
J. Xu, L. Chen, and S. Ren, ``Online learning for offloading and autoscaling in energy harvesting mobile edge computing,'' {\em IEEE Trans. Cognitive Commun. Netw.}, vol. 3, no. 3, pp. 361--373, Sep. 2017.


\bibitem{Li15}
H. Li, J. Xu, R. Zhang, and S. Cui, ``A general utility optimization framework for energy harvesting based wireless communications,'' {\em IEEE Commun. Mag.}, vol. 53, no. 4, pp. 79--85, Apr. 2015.

\bibitem{Rahbar15}
K. Rahbar, J. Xu, and R. Zhang, ``Real-time energy storage management for renewable integration in microgrid: An offline optimization approach,'' {\em IEEE Trans. Smart Grid}, vol. 6, no. 1, pp. 124--134. Jan. 2015.








\bibitem{Boyd2004} S.~Boyd and L.~Vandenberghe, {\em Convex Optimization}, Cambridge, U.K.: Cambridge Univ. Press, 2004.
%
%

\end{thebibliography}


\end{document}